\documentclass[a4paper,11pt,twoside]{article}

\usepackage{amsmath}
\usepackage{amssymb}
\usepackage{epsfig}

\newcommand{\GeV}{\,\mathrm{GeV}}

\newcommand{\eV}{\,\mathrm{eV}}
\newcommand{\vev}[1]{\left\langle #1\right\rangle}

\DeclareMathOperator{\re}{Re}
\newcommand{\ord}[1]{\mathcal{O}\left( #1 \right)}

\newcommand{\capdef}{}
\newcommand{\mycaption}[2][\capdef]{\renewcommand{\capdef}{#2}%
        \caption[#1]{{\itshape #2}}} 
\makeatletter
\renewcommand{\fnum@table}{\textbf{\tablename~\thetable}}
\renewcommand{\fnum@figure}{\textbf{\figurename~\thefigure}}
\makeatother

\newlength{\myem}
\newcommand{\sep}[1]{#1}
\newcounter{mysubequation}[equation]
\renewcommand{\themysubequation}{\alph{mysubequation}}
\newcommand{\mytag}{\stepcounter{mysubequation}%
\tag{\theequation\protect\sep{\themysubequation}}}
\newcommand{\globallabel}[1]{\refstepcounter{equation}\label{#1}}

\makeatletter
\renewcommand{\section}{\@startsection{section}{1}{0em}%
	{-3.5ex \@plus -1ex \@minus -.2ex}%
	{2.3ex \@plus.2ex}%
	{\normalfont\large\bfseries}}
\renewcommand{\subsection}{\@startsection{subsection}{2}{0em}%
	{-3.25ex\@plus -1ex \@minus -.2ex}%
	{1.5ex \@plus .2ex}%
	{\normalfont\bfseries}}
\renewcommand{\subsubsection}%
	{\@startsection{subsubsection}{3}{0em}%
	{-3.25ex\@plus -1ex \@minus -.2ex}%
	{1.5ex \@plus .2ex}%
	{\normalfont\itshape}}
\makeatother

\newcommand{\dm}[1]{{\Delta m^2_{#1}}}
\newcommand{\reu}{{\nu_e\rightarrow\nu_\mu}}
\newcommand{\reub}{{\bar{\nu}_e\rightarrow\bar{\nu}_\mu}}
\newcommand{\Nup}{{N_{\mu^+}}}
\newcommand{\Num}{{N_{\mu^-}}}
\newcommand{\nup}{{n(\nu_\mu)}}
\newcommand{\num}{{n(\bar{\nu}_\mu)}}
\newcommand{\nump}{{n'(\bar{\nu}_\mu)}}
\newcommand{\emax}{{E_{\text{max}}}}
\newcommand{\emin}{{E_{\text{min}}}}
\newcommand{\NKT}{{N_{\text{kT}}}}
\newcommand{\aCP}{{a_{\text{CP}}}}
\newcommand{\aCPd}{{a_{\text{CP}}(\delta=\pi/2)}}
\newcommand{\dasy}{{\delta_{\text{syst}}a_{\text{CP}}}}
\newcommand{\dast}{{\delta_{\text{stat}}a_{\text{CP}}}}
\newcommand{\PCPV}{
\begin{picture}(22,10)
\put(8,-2){\line(2,1){12}}
\put(0,0){$P_{CP}$}
\end{picture}}
\newcommand{\PCPC}{
\begin{picture}(22,10)
\put(0,0){$P_{CP}$}
\end{picture}}

\newcommand{\OX}{Department of Physics, Theoretical Physics,
University of Oxford, Oxford OX1\hspace{0.2em}3NP, UK}

\newcommand{\preprintdate}{September 1999}
\newcommand{\preprintnumber}{OUTP--99--46P}

\newcommand{\titletext}{Measuring CP-violation with a neutrino factory}
\newcommand{\authortext}{\large Andrea Romanino
\medskip\\\em\normalsize \OX} \newcommand{\abstracttext}{We discuss
the prospects of observing leptonic CP-violation at a neutrino factory
in the context of the standard three neutrino scenario.  If the large
mixing angle MSW effect turned out to account for the solar
neutrino deficit, we show that observing an asymmetry between the
$\reu$ and $\reub$ oscillation probabilities would represent an
exciting experimental challenge.  We determine the portion of the
parameter space where an evidence could be found as a function of the
intensity of the muon source and of the detector size for different
baselines. We discuss the consequent requirements on a neutrino
factory. We address the problems associated with asymmetries induced
by the experimental apparatus and by matter effects. Finally, we
emphasize the importance of measuring two CP-conjugated channels in
order to precisely determine $\theta_{13}$.}

\title{
\normalsize
\begin{tabular}[t]{l}\end{tabular}
\hspace*{\fill}
\begin{tabular}[t]{l}\preprintnumber\\\preprintdate\end{tabular}
\vspace{3\baselineskip}\\\Large\bfseries\titletext\bigskip}
\author{\begin{minipage}[t]{0.8\textwidth}
\normalsize\centering\authortext
\end{minipage}}
\date{}

\begin{document}

\bigskip
\maketitle
\begin{abstract}
\normalsize\noindent\abstracttext
\end{abstract}
\normalsize\vspace{\baselineskip}

\section{Introduction}
\label{sec:intro}

\noindent
An increasing evidence for the existence of neutrino oscillations has
been gathered by several experiments. The standard theoretical
interpretations of the data requires the neutrino flavour eigenstates
to be superpositions of three or more mass eigenstates. This involves
the possibility of a violation of CP associated with the physical
phases of the mixing matrix. 

An evidence of CP-violation has been provided in the quark sector by
different experiments, that have suggested a value of the CKM phase
close to maximal. It would be extremely interesting to be able to
investigate the CP-violation issue also in the lepton
sector. Unfortunately, the upcoming long-baseline experiments will not
help in this respect since they are not conceived for the purpose of
comparing CP-conjugated transition. Note, however, that they could be
affected by CP-violation in a relevant way. The oscillation
probabilities they measure do in fact contain a CP-violating part. A
detailed analysis~\cite{DFLR} has shown that the latter can constitute
a relevant part of the full oscillation probability in the
$\nu_e\leftrightarrow\nu_\mu$ channel and could therefore affect the
measurement of the mixing angle $\theta_{13}$.

An experimental study of CP-violation in neutrino oscillations would
require a further generation of experiments offering the possibility
of measuring oscillation probabilities in two CP-conjugated
channels. Such possibility has been made concrete by the proposal of
using the very intense muon sources that are currently being designed
as part of the muon collider projects to produce an high-intensity
neutrino beam~\cite{geer:98a,derujula:98a,BGW}. As a pure source of
both neutrinos and antineutrinos with well known initial flux, such a
``neutrino factory'' would be an ideal framework for studying leptonic
CP-violation.  The possibility of exploiting a neutrino factory for a
measurement of CP-violation has been first considered
in~\cite{derujula:98a}, where the capabilities of a reference set-up
have been studied in different points of the neutrino parameter
space. An update of that analysis can be found in~\cite{DGHR}. The
purpose of this paper is to explore the full parameter space and to
show in which part of it the detection of a CP-violating effect would
be possible. The portion of the parameter space that could be covered
will be determined as a function of the intensity of the muon source
and of the size and efficiency of the detector for two different
baselines. This will make clear which values of the factory parameters
are necessary in order to cover a given portion of the parameter
space. We address these issues in the context of three neutrino
oscillations. Analysis of CP-violation in four neutrino oscillations
can be found in~\cite{barger:99a,DFLR,DGHR,KMT}.

The paper is organized as follows. In Section~\ref{sec:framework} we
introduce the usual notations for the neutrino masses and mixings and
we describe the experimental set-up. In Section~\ref{sec:measuring} we
discuss the problems involved with a realistic measurement of
CP-violation. In particular we describe how we cope with the asymmetry
induced by the experimental apparatus and by matter effects. The
systematic and statistical errors associated with the measurement are
also discussed. In Section~\ref{sec:requirements} we determine the
optimal values of some parameters of the experiment and we present the
results about its capabilities for different experimental
configurations. In Section~\ref{sec:conc} we discuss the results and
we conclude.

\section{Framework}
\label{sec:framework}

\subsection{Masses, mixings and probabilities}
\label{subsec:frame}

\noindent
In this paper we consider the possibility of measuring a violation of
CP in the leptonic sector by means of a neutrino
factory~\cite{geer:98a}.
We address this issue in the context of three neutrino
oscillations. The leptonic charged current is then given by
\begin{equation}
\label{cc}
\bar{e_i}\gamma^\mu \nu_{e_i} = \bar{e_i}\gamma^\mu U_{ij}\nu_j,
\end{equation}
where $U_{ij}$ is the $3\times3$ unitary lepton mixing matrix and
$e_i$, $\nu_i$, $i=1\ldots 3$ are the left-handed charged lepton and
neutrino mass eigenstates. By convention we order the neutrino mass
eigenstates in such a way that $m_{\nu_1}<m_{\nu_2}$ and
$\dm{21}<|\dm{31}|, |\dm{32}|$, where we denote as usual
$\dm{ij}=m^2_{\nu_i}-m^2_{\nu_j}$. The common sign of $\dm{31}$ and
$\dm{32}$ is not determined at present.

We use the standard parameterization for $U$, 
\begin{equation} 
U= 
\begin{pmatrix} 
  c_{12}c_{13} & s_{12}c_{13} & s_{13}e^{-i\delta} \\
  -s_{12}c_{23} -c_{12}s_{23}s_{13}e^{i\delta} & 
  c_{12}c_{23}
  -s_{12}s_{23}s_{13}e^{i\delta} & s_{23}c_{13} \\
  s_{12}s_{23} -c_{12}c_{23}s_{13}e^{i\delta} & 
  -c_{12}s_{23}
  -s_{12}c_{23}s_{13}e^{i\delta} & c_{23}c_{13}
\end{pmatrix},
\label{par} 
\end{equation} 
where possible Majorana phases are omitted because they do not enter
the oscillation formulae in the safe approximation in which lepton
number violating oscillations are neglected\footnote{We prefer to
think of three Majorana neutrinos but in the mentioned approximation
what follows apply to the case of three purely Dirac neutrinos as
well.}.  In that approximation, the
$\nu_{e_i}\rightarrow\nu_{e_i}$,
$\bar{\nu}_{e_i}\rightarrow\bar{\nu}_{e_j}$ oscillation probabilities
in vacuum are \globallabel{Pne}
\begin{align}
P(\nu_{e_i}\rightarrow\nu_{e_j}) &=
\PCPC(\nu_{e_i}\rightarrow\nu_{e_j}) + \PCPV(\nu_{e_i}\rightarrow\nu_{e_j})
\mytag \\
P(\bar{\nu}_{e_i}\rightarrow\bar{\nu}_{e_j}) &=
\PCPC(\nu_{e_i}\rightarrow\nu_{e_j}) - \PCPV(\nu_{e_i}\rightarrow\nu_{e_j}),
\mytag
\end{align}
where 
\globallabel{3n}
\begin{align}
\PCPC(\nu_{e_i}\rightarrow\nu_{e_j}) &= \delta_{ij} -4\re J^{ji}_{12}
\sin^2\Delta_{12} -4\re J^{ji}_{23} \sin^2\Delta_{23} -4\re J^{ji}_{31} 
\sin^2\Delta_{31}, \mytag \\
\PCPV(\nu_{e_i}\rightarrow\nu_{e_j})
& = -8\sigma_{ij} J \sin\Delta_{12} 
\sin\Delta_{23} \sin\Delta_{31}, \mytag  
\end{align}
$J^{e_i e_j}_{kh} \equiv U_{e_i\nu_k} U_{\nu_k e_j}^\dagger
U_{e_j\nu_h} U_{\nu_h e_i}^\dagger$, $\Delta_{kh}\equiv \Delta m^2_{kh}
L/4E$, $\sigma_{ij}\equiv \sum_k \varepsilon_{ijk}$ and 
\begin{equation}
8 J = \cos\theta_{13} \sin(2\theta_{13}) \sin(2\theta_{12})
\sin(2\theta_{23}) \sin\delta.
\label{Jct}
\end{equation}

The ranges we use for the neutrino masses and mixings are those
obtained by the standard fits of the atmospheric and solar neutrino
data~\cite{fukuda:98b,GHPV,fukuda:99a,barger:98a} taking into account
the constraints given by the CHOOZ
experiment~\cite{apollonio:97a}. Among the three possible standard
solutions of the solar neutrino problem we only consider in detail the
``large angle'' one. In the ``small angle'' solution, as well as in
the ``vacuum'' solution, CP-violation effects are in fact too small to
be detectable, since they are suppressed either by the small value of
$\theta_{12}$ (small angle solution) or by the small value of
$\dm{32}$ (vacuum solution).

\subsection{Experimental set-up}
\label{subsec:experimental}

\noindent
The possibility of using the straight section of a high intensity muon
storage ring as a neutrino factory has been emphasized
in~\cite{geer:98a,derujula:98a}. Being a pure and high intensity
source of both neutrinos and antineutrinos with well known initial
flux, a neutrino factory would give the possibility of comparing the
oscillation probabilities in two CP-conjugated channels:
$\nu_{e_i}\rightarrow\nu_{e_j}$/$\bar{\nu}_{e_i}\rightarrow\bar{\nu}_{e_j}$
($i\neq j$).  Therefore, it would be the ideal framework for studying
leptonic CP-violation.  Among the possible channels made available by
a neutrino factory, the most suitable for studying CP-violation are
$\reu$/$\reub$ and the T-conjugated
$\nu_\mu\rightarrow\nu_e$/$\bar{\nu}_\mu\rightarrow\bar{\nu}_e$. In
fact, in these channels the CP-violating part of the oscillation
probability is not submerged by the CP-conserving part (as it is for
the $\nu_\mu\leftrightarrow\nu_\tau$ channels), so that large
asymmetries between the CP-conjugated channels can
arise~\cite{DFLR}. We consider here in particular the $\reu$/$\reub$
channels, since telling $\mu^+$ from $\mu^-$ in a large high-density
detector is easier than telling $e^+$ from $e^-$~\cite{derujula:98a}.

The $\reu$ ($\reub$) oscillation probability can be measured as
follows: the electron neutrinos (antineutrinos) are produced by the decay of
$\Nup$ positive muons ($\Num$ negative muons) in the straight section
of the storage ring pointing to the detector; then the $\nu_\mu$
($\bar{\nu}_\mu$) produced by the oscillation can be detected by their
charged current interactions in the detector. The number $\nup$
($\num$) of ``observed'' $\nu_\mu$ ($\bar{\nu}_\mu$) is then given by 
\globallabel{nunum}
\begin{align}
\nup &= \Nup \NKT \frac{10^9
N_A}{m^2_\mu\pi}\frac{E^3_\mu}{L^2}\int^\emax_\emin
f(E)P^m_\reu(E)\frac{dE}{E_\mu} \mytag \\
\num &= \Num \NKT \frac{10^9
N_A}{m^2_\mu\pi}\frac{E^3_\mu}{L^2}\int^\emax_\emin
\bar{f}(E)P^m_\reub(E)\frac{dE}{E_\mu}, \mytag 
\end{align}
where 
\globallabel{fdef}
\begin{align}
f(E) &=g_{\nu_e}(E/E_\mu)
(\sigma_{\nu_\mu}(E)/E_\mu)\epsilon_{\mu^-}(E), \mytag \\ \bar{f}(E)
&=g_{\bar{\nu}_e}(E/E_\mu)
(\sigma_{\bar{\nu}_\mu}(E)/E_\mu)\epsilon_{\mu^+}(E) \mytag
\end{align}
are weight functions taking into account the energy spectrum
(normalized to 1) of the electron neutrinos (antineutrinos) produced
in the $\mu^+$ ($\mu^-$) decay, $g_{\nu_e (\bar{\nu}_e)}(E/E_\mu)$,
the charged current cross section per nucleon, $\sigma_{\nu_\mu
(\bar{\nu}_\mu)}(E)$, and the efficiency for the detection of $\mu^-$
($\mu^+$), $\epsilon_{\mu^- (\mu^+)}(E)$\footnote{We neglect here the
finite resolution of the detector.}. $P^m_\reu(E)$ ($P^m_\reub(E)$) is
the oscillation probability for neutrinos travelling inside the earth
taking into account matter effects. Finally $\NKT$ is the size of the
detector in kilotons and $N_{\mu^{\pm}}$ is the number of ``useful''
$\mu^{\pm}$ decays, namely the number of decays occurring in the
straight section of the storage ring pointing to the detector. In the
numerical calculations we use \globallabel{numeric}
\begin{gather}
g_{\nu_e}(x) = g_{\bar{\nu}_e}(x) =12x^2(1-x), \mytag \\
\sigma_{\nu_\mu}(E) = 0.67\cdot 10^{-38}E\,\text{cm}^2/\text{GeV},
\qquad \sigma_{\bar{\nu}_\mu}(E) = 0.34\cdot
10^{-38}E\,\text{cm}^2/\text{GeV}, \mytag \\ \text{and}\quad
\epsilon_{\mu^\pm}(E) = 30\%\quad\text{for $E>5\GeV$}, \mytag
\end{gather}
so that $f(E)/\bar{f}(E)=2$ (independently of the energy). We also
apply a lower cut $\emin=5\GeV$ on the neutrino energies in order to
have a good detection efficiency in all the energy range\footnote{The
numerical results we obtain agree with those obtained in ~\cite{DGHR}
if the efficiency is set to 100\% and no lower cut is used.}.

\section{Measuring CP-violation}
\label{sec:measuring}

\subsection{Experimental apparatus and matter asymmetries}
\label{subsec:fake}

\noindent
The asymmetry between the number of $\nu_\mu$ and
$\bar{\nu}_\mu$ events seen in the detector, normalized to the initial
number of muon decays is given by
\begin{equation}
\label{atotmea}
\hat{a}_{\text{tot}} = \frac{\nup/\Nup-\num/\Num}{\nup/\Nup+\num/\Num}.
\end{equation}
An ``intrinsic'' leptonic CP-violation associated with a non-vanishing
phase $\delta$ in~(\ref{par}) would contribute to give a non-vanishing
$\hat{a}_{\text{tot}}$. However, even if there were not any intrinsic
CP-violation, $\hat{a}_{\text{tot}}$ would still be non-vanishing due
to the CP-asymmetry of the experimental apparatus, $f\neq\bar{f}$
in~(\ref{nunum}), and due to matter effects, that affect neutrinos and
antineutrinos in different ways. In other words,
$\hat{a}_{\text{tot}}$ is an estimator of
\begin{equation}
\label{atot}
a_{\text{tot}} = \frac{\int P^m_\reu(E)f(E)\,dE - \int
P^m_\reub(E)\bar{f}(E)\,dE}{\int P^m_\reu(E)f(E)\,dE + \int
P^m_\reub(E)\bar{f}(E)\,dE},
\end{equation}
that can be non-vanishing even when $\delta=0$ due to
$f\neq\bar{f}$ and $P^m_\reu(E,L)\neq P^m_\reub(E,L)$.

Getting rid of the experimental apparatus asymmetry is
easy. It is sufficient to weight e.g.\ each $\bar{\nu}$ event with
$f(E)/\bar{f}(E)$, where $E$ is the energy of the event, namely to
replace $\num$ in~(\ref{atotmea}) with
\begin{equation}
\label{weight}
\nump\equiv\sum^\num_{i=1} \frac{f(E_i)}{\bar{f}(E_i)},
\end{equation}
where $E_i$ is the energy of the $i$-th $\bar{\nu}_\mu$ event. In other
words, instead of $\hat{a}_{\text{tot}}$ we consider
\begin{equation}
\label{aCPmmea}
\hat{a}_{\text{CP+m}} = \frac{\nup/\Nup-\nump/\Num}{\nup/\Nup+\nump/\Num},
\end{equation}
that estimates a quantity independent of the experimental
CP-violation, 
\begin{equation}
\label{aCPm}
a_{\text{CP+m}} = \frac{\int (P^m_\reu(E) -
P^m_\reub(E))f(E)\,dE}{\int (P^m_\reu(E) +
P^m_\reub(E))f(E)\,dE}.
\end{equation}
We remark, however, that the subtraction of the experimental asymmetry
introduces an unavoidable systematic error due to the uncertainties in the
knowledge of the flux, cross section and efficiency ratios in
$f(E)/\bar{f}(E)$.

In our numerical calculations $\nump$ is simply given by
\begin{equation}
\label{num}
\nump=2\num,
\end{equation}
since $f(E)/\bar{f}(E)=2$ independently of the neutrino energy $E$. In
this case, using~(\ref{aCPmmea}) is equivalent to using a different
normalization for $\nup$ and $\num$ in~(\ref{atotmea}). However, if
$f(E)/\bar{f}(E)$ had a small dependence on the energy due e.g.\ to
$\epsilon_{\mu^-}(E)/\epsilon_{\mu^+}(E)$, using~(\ref{aCPmmea}) would
allow a more accurate subtraction of the experimental asymmetry.

The quantity $\hat{a}_{\text{CP+m}}$ in~(\ref{aCPmmea}) still contains
a pollution due to the matter asymmetry.  In principle the best
attitude towards such a pollution would be trying to experimentally
disentangle matter from genuine CP-violation effects. A detailed study
of the feasibility of this possibility is beyond the aim of this
paper. Here we would rather cope with matter effects by calculating
them~\cite{DFLR}. An experimental confirmation of the reliability of
such a theoretical calculation, besides being important in itself,
would be welcome in order to support this method. Other theoretical
approaches to the issue of matter effects can be found
in~\cite{arafune:97a}.

First of all, we define the matter asymmetry in absence of
CP-violation as
\begin{equation}
\label{am}
a_{\text{m}} = \frac{\int (P^{m,\mathbf{\delta=0}}_\reu(E) -
P^{m,\mathbf{\delta=0}}_\reub(E))f(E)\,dE}{\int
(P^{m,\mathbf{\delta=0}}_\reu(E) +
P^{m,\mathbf{\delta=0}}_\reub(E))f(E)\,dE},
\end{equation}
where $P^{m,\mathbf{\delta=0}}_\reu$, $P^{m,\mathbf{\delta=0}}_\reub$
are the oscillation probabilities calculated for $\delta=0$ but taking
into account matter effects. Then, we define the CP-asymmetry
in absence of matter effects as 
\begin{equation}
\label{aCP}
a_{\text{CP}} = \frac{\int (P_\reu(E) - P_\reub(E))f(E)\,dE}{\int
(P_\reu(E) + P_\reub(E))f(E)\,dE} =
\frac{\int \PCPV(E) f(E)\,dE}{\int \PCPC(E)
f(E)\,dE}, 
\end{equation}
where $P_\reu$, $P_\reub$ are the oscillation probabilities computed
in vacuum but taking into account CP-violation effects. Now, despite
$a_{\text{CP+m}}$ is not simply given by the sum of $a_{\text{CP}}$
and $a_{\text{m}}$, it turns out that for practical purposes the
relation
\begin{equation}
\label{subtr}
a_{\text{CP+m}} \simeq a_{\text{CP}} + a_{\text{m}}
\end{equation}
is a very good approximation if $a_{\text{m}}$ is not too
large\footnote{This is essentially because the corrections
to~(\ref{subtr}) arise only at second order in the asymmetries and is
confirmed by a numerical analysis~\cite{DFLR}.}. In any case, the
error one makes recovering $a_{\text{CP}}$ through~(\ref{subtr}),
$a_{\text{CP}} = a_{\text{CP+m}}-a_{\text{m}}$, is smaller than the
uncertainties on $a_{\text{m}}$ itself. The latter therefore
constitute the main source of systematic error from matter effects.
In conclusion, as a measure of CP-violation we use the quantity
\begin{equation}
\label{aCPmea}
\hat{a}_{\text{CP}} =
\frac{\nup/\Nup-\nump/\Num}{\nup/\Nup+\nump/\Num}-a_{\text{m}}.
\end{equation}
Being $\hat{a}_{\text{CP}}$ an estimator of $a_{\text{CP}}$
in~(\ref{aCP}), measuring a non-vanishing value of
$\hat{a}_{\text{CP}}$ would indicate a genuine leptonic CP-violation.

\subsection{Systematic and statistical errors}
\label{subsec:errors}

\noindent
In this Section we discuss the systematic and statistical errors
involved in a measurement of $a_{\text{CP}}$ through
$\hat{a}_{\text{CP}}$.  There are different sources of systematic
errors in $\hat{a}_{\text{CP}}$. One is given by the uncertainties on
the experimental asymmetries, $g_{\nu_e}/g_{\bar{\nu}_e}$,
$\sigma_{\nu_{\mu}}/\sigma_{\bar{\nu}_{\mu}}$ and
$\epsilon_{\mu^-}/\epsilon_{\mu^+}$, that enter $\hat{a}_{\text{CP}}$
through $f/\bar{f}$ in~(\ref{weight}). Another one is given by the
uncertainties on the matter asymmetry, in turn due to the
uncertainties on the neutrino mass and mixing parameters we use to
calculate it. Here we concentrate on the latter, assuming that the
former will be made small enough by experimental determinations of
$f/\bar{f}$.

Whatever is the strategy used towards the matter asymmetry, it is
clear that the smaller the matter effects are the cleaner a
possible measurement of a genuine violation of CP would be. Since the
matter asymmetry grows with the baseline faster than the CP-asymmetry,
$a_{\text{m}}$ constitutes the main limitation to the length of the
baseline of an experiment aiming at measuring a violation of CP. In
order to illustrate and make quantitative the latter point we can use
the approximation one gets for $P^m_\reu$, $P^m_\reub$ in the
limit $\dm{21}=0$:
\globallabel{crude}
\begin{align}
P^m_\reu &= \sin^2\theta_{23}\sin^2
2\theta_{13}\frac{\sin^2(\Delta_{32}c^{1/2}_+)}{c_+} \mytag
\\
P^m_\reub &= \sin^2\theta_{23}\sin^2
2\theta_{13}\frac{\sin^2(\Delta_{32}c^{1/2}_-)}{c_-}, \mytag
\end{align}
where 
\begin{equation}
c_\pm=(1\mp x)^2\pm4x\sin^2\theta_{13} \simeq (1\mp x)^2, \quad
x=\frac{2EV}{\dm{32}}
\end{equation}
and $V$ is the matter induced potential. We then get 
\begin{equation}
\label{am0}
a_{\text{m}}=\frac{\sin^2(\Delta_{32}c^{1/2}_+)/c_+ -
\sin^2(\Delta_{32}c^{1/2}_-)/c_-}
{\sin^2(\Delta_{32}c^{1/2}_+)/c_+ 
+\sin^2(\Delta_{32}c^{1/2}_-)/c_-} \simeq
\frac{V\dm{32}}{12}\frac{L^2}{E}(1-2\sin^2\theta_{13}) 
\end{equation}
for the matter asymmetry before integration on the energy range,
or
\begin{equation}
\label{amapp}
a_{\text{m}}\simeq 0.7\cdot 10^{-6}\frac{L^2(\text{km})}{E(\GeV)}
\frac{\dm{32}}{3\cdot 10^{-3}\eV^2}.
\end{equation}
The expansion in~(\ref{am0}) holds when $L\lesssim\pi/V$ and
$\Delta_{32}\lesssim\pi/2$. When these conditions are not fulfilled
the matter asymmetry is large or quickly oscillating and therefore out
of control. However, we stress that the main corrections
to~(\ref{amapp}) come from the approximation~(\ref{crude}),
particularly rough when $\theta_{13}$ is very small, rather than from
the expansion in~(\ref{am0}). Nonetheless, Eq.~(\ref{amapp}) is
sufficient for our illustrative purposes (exact formulae are used in
all numerical calculations). When compared with the behavior of
$a_{\text{CP}}$ with $L$ and $E$,
\begin{equation}
\label{aCPLE}
a_{\text{CP}}\propto\sin\Delta_{21}\propto\frac{L}{E},
\end{equation}
Eq.~(\ref{amapp}) shows that $|a_{\text{m}}/a_{\text{CP}}|$ grows with
the baseline length $L$ (as far as $L\lesssim\pi/V$ and
$\Delta_{32}\lesssim\pi/2$). Very long baselines are therefore more
suitable for the study of matter effects than for CP-violation. We
will come back to this point in Section~\ref{sec:requirements}. 

Before describing into detail how we work out the systematic error due
to matter effects, we remark that Eq.~(\ref{amapp}), besides
illustrating the linear dependence of $a_{\text{m}}$ on $\dm{32}$,
makes also explicit its dependence on the sign of $\dm{32}$. The
knowledge of that sign is therefore essential in order to properly
subtract matter effects and will be assumed here. A detailed analysis
of how that sign could be determined through a measurement of matter
effects is beyond the scope of this paper.

We estimate the systematic error $\dasy$ on $a_{\text{CP}}$ due to the
uncertainty on $a_{\text{m}}$ by scanning the ranges of the
parameters on which $a_{\text{m}}$ depends. $\dasy$ is then obtained as the
half-width of the range of values assumed by $a_{\text{m}}$. As for
the $\dm{21}$, $\theta_{12}$, $\theta_{23}$ ranges, less important in the
determination of $\dasy$, we assume, to be conservative, that they
will not differ very much from the present
ones~\cite{fukuda:98b,GHPV,fukuda:99a,barger:98a}:
\begin{equation}
\label{ranges}
\dm{21}=(0.2 - 3.0)\cdot 10^{-4}\eV^2, \quad \sin^2 
2\theta_{12} = (0.6 - 0.96), \quad \sin^2
2\theta_{23}> 0.8
\end{equation}
(all at 99\% CL; the $\dm{21}$, $\theta_{12}$ values correspond of
course to the large mixing angle solution). It is less clear which
ranges should be used for the more crucial $\dm{32}$ and
$\theta_{13}$. The next generation of long-baseline experiments will
be able to provide information on the latter parameters and will
reduce the uncertainties on them. However, it has been shown
in~\cite{DFLR} that, for favourable values of $\dm{21}$ in the large
angle range in~(\ref{ranges}) and for values of $\dm{32}$ and
$\theta_{13}$ within the sensitivity of an experiment like e.g.\
MINOS, the CP-violating part of the oscillation probability could
constitute up to 30--40\% of the total oscillation probability. This in
turn means that, unless some compelling evidence excluding the large
angle solution is found, a systematic uncertainty up to 30--40\% could
affect a $\sin^2 2\theta_{13}$ determination made measuring a single
oscillation probability. A precise determination of $\theta_{13}$
would on the contrary require an experiment able to measure both the
$\nu_e\leftrightarrow\nu_\mu$ and
$\bar{\nu}_e\leftrightarrow\bar{\nu}_\mu$ oscillation probabilities,
namely able to measure CP-violation. A neutrino factory would provide
this possibility. Once both $P_\reu$ and $P_\reub$ were measured, the
CP-conserving part of the probability could be recovered as $\PCPC =
(P_\reu+P_\reub)/2$. This would not completely remove the ambiguities
associated to the phase $\delta$, since $\PCPC$ depends on $\delta$ as
well. An estimate of $\delta$ through $\aCP$ would therefore be also
welcome in order to get rid of such ambiguities. Here, for the purpose
of estimating $\dasy$ we will assume that a determination of $\dm{32}$
and $\sin^2 2\theta_{13}$ with a precision higher than 20\% will
become available.

The discussion of statistical uncertainties is straightforward. In
order to exclude the possibility that a measurement of a non-vanishing
$\hat{a}_{\text{CP}}$ is due to a statistical fluctuation, the measured
value  must be larger than $n_\sigma\dast$, where
$\dast$ is the ``$1\sigma$'' statistical error on $\aCP$ in absence of
CP-violation and $n_\sigma$ is the number of standard deviations we
require. Since in absence of CP-violation the expectations of $\nup/\Nup$ and
$\nump/\Num$ are equal, we get
\begin{equation}
\label{stat1}
\dast = \left(\frac{1}{4\vev{\nup}}+\frac{1}{4\vev{\num}}\right)^{1/2},
\end{equation}
where $\vev{\nup}$ and $\vev{\num}$ are the expected number of
$\nu_\mu$ and $\bar{\nu}_\mu$ interactions seen in the detector. When
deciding whether it would be possible to measure a violation of CP in
a given point of parameter space, we will assume as usual that the
measured value of $\hat{a}_{\text{CP}}$ coincides with the value of
$\aCP$ expected in that point.

Finally, let us shortly discuss the contribution of the background to
the statistical error. According to the present estimates, the main
source of background is due to charm production in the charged and
neutral current
neutrino interactions in the detector~\cite{CDG}. For example, the
charged current background is due to:
\begin{equation}
\label{back}
\mu^- \rightarrow \nu_\mu \xrightarrow{\text{no osc.}} \nu_\mu
\xrightarrow{\text{CC int.}} (\mu^-)_{\text{lost}} + c
\xrightarrow{\text{$c$-decay}} (\mu^+)_{\text{found}}.
\end{equation}
This background only affects the signal corresponding to the $\reub$
oscillation:
\begin{equation}
\label{signal}
\mu^- \rightarrow \bar{\nu}_e \xrightarrow{\text{osc.}}\bar{\nu}_\mu
\xrightarrow{\text{CC int.}} \mu^+.
\end{equation}
The background subtraction corresponds to a replacement 
\begin{equation}
\label{back2}
\num\rightarrow \num-(\num)_{\text{back}}
\end{equation}
in Eq.~(\ref{aCPmea}). Such a subtraction introduces a further source
of statistical error. Using the estimate~\cite{CDG}
\begin{equation}
\label{estimate}
(\num)_{\text{back}}\sim 10^{-5}\nump_{P=1},
\end{equation}
where $\num_{P=1}$ stands for the number of $\bar{\nu}_\mu$
interactions that would be seen if all the initial $\bar{\nu}_e$
oscillated into $\bar{\nu}_\mu$, we get the complete expression for
the statistical error we will use:
\begin{equation}
\label{stat}
\dast = \left(\frac{1}{4\vev{\nup}}+\frac{1}{4\vev{\num}} +
\frac{10^{-5}\vev{\num_{P=1}}}{4\vev{\num}^2}\right)^{1/2}. 
\end{equation}
Note, however, that in a long-baseline experiment with
$\sin^2\Delta_{32}=\ord{1}$ the background contribution in~(\ref{stat})
is not negligible only for $\sin^2 2\theta_{13}\lesssim
10^{-5}$.

\section{Capability of the neutrino factory}
\label{sec:requirements}

\noindent
In this Section we show in which part of the parameter space an
evidence of leptonic CP-violation could be obtained as a function of
the relevant experimental parameters. In order to obtain an evidence
for a non-vanishing 
\begin{equation}
\label{aCP2}
a_{\text{CP}} = \frac{\int (P_\reu(E) - P_\reub(E))f(E)\,dE}{\int
(P_\reu(E) + P_\reub(E))f(E)\,dE} =
\frac{\int \PCPV(E)f(E)\,dE}
{\int \PCPC(E)f(E)\,dE},
\tag{\ref{aCP}}
\end{equation}
a value of $\hat{a}_{\text{CP}}$
larger than the systematic and statistical errors on it should be measured:
\begin{equation}
\label{test}
|\hat{a}_{\text{CP}}| > \dasy, \qquad 
|\hat{a}_{\text{CP}}| > n_\sigma \dast.
\end{equation}
In the previous equation, $\dasy$ is calculated as described in
Section~\ref{subsec:errors}, 
$\dast$ is the $1\sigma$ statistical error given by~(\ref{stat}) and
$n_\sigma$ is the number of standard deviations corresponding to the
CL we want to achieve (we will use $n_\sigma = 5$ as a reference value).

The experimental parameters we consider are
\begin{itemize}
\item
the size of the detector in kT, $\NKT$;
\item
the total number of useful muon decays $N_\mu=\Nup+\Num$;
\item
the ratio of $\mu^+$ versus $\mu^-$ decays $\Nup/\Num$ for a given
total number of muon decays $N_\mu=\Nup+\Num$;
\item
the energy of the stored muons $E_\mu$ and the lower and upper cuts
on the neutrino energies, $\emin$ and $\emax$;
\item
the baseline $L$.
\end{itemize}

First, we briefly identify a convenient choice of $\Nup/\Num$ and
$E_\mu$, $\emin$, $\emax$. Then, using that choice, we show which is the
portion of parameter space where the asymmetry could be
measurable as a function of the product
$N_\mu\NKT$ and for two choices of the baseline, $L=732\,\text{km}$
and $L=3000\,\text{km}$. 
\medskip

The dependence of the statistical error on $\Nup$ and $\Num$ is
simple. Neglecting the small background contribution and using
Eqs.~(\ref{numeric}) we find
\begin{equation}
\label{nratio}
\dast \propto
\left(\frac{1}{\Nup}+\frac{1}{\Num}
\frac{\sigma_{\nu_\mu}}{\sigma_{\bar{\nu}_\mu}}\right)^{1/2},
\end{equation}
where $\sigma_{\nu_\mu}/\sigma_{\bar{\nu}_\mu}\simeq 2$. When
$\Nup=\Num$ the $\bar{\nu}_\mu$ events are less than the $\nu_\mu$ events
because of the smaller $\bar{\nu}_\mu$ cross section. Therefore, their
contribution to the statistical error is larger. For a given total
number of available muon decays, $N_\mu=\Nup+\Num$, the statistical
error can be made smaller by choosing a ratio $\Nup/\Num$ smaller than
one. Here we therefore use the $\Nup/\Num$ ratio
obtained by minimizing ~(\ref{nratio}), namely \globallabel{Nopt}
\begin{align}
\Nup &= \frac{\sqrt{\sigma_{\bar{\nu}_\mu}}}
{\sqrt{\sigma_{\nu_\mu}}+\sqrt{\sigma_{\bar{\nu}_\mu}}} N_\mu \simeq
0.4 \,N_\mu \mytag \\ 
\Num &= \frac{\sqrt{\sigma_{\nu_\mu}}}
{\sqrt{\sigma_{\nu_\mu}}+\sqrt{\sigma_{\bar{\nu}_\mu}}} N_\mu \simeq
0.6 \,N_\mu. 
\mytag
\end{align}
Unfortunately, using~(\ref{Nopt}) instead of $\Nup=\Num$ only improves
$\dast$ by a few percents. 

Let us now discuss the energy parameters. First of all, we use
$\emin=5\GeV$ as a lower cut on the neutrino energies in order to have
an efficient muon detection in all the energy range. An upper cut is
also necessary since $\aCP/\dast \propto1/\sqrt{\emax}$ when $\emax$
is sufficiently high~\cite{DGHR}. The upper cut will of course be
lower than the muon energy in the storage ring. The low energy
neutrino rates do not strongly depend on the value of
$E_\mu\geq\emax$, as far as $E_\mu$ is in the range presently
discussed $10\GeV < E_\mu < 50\GeV$. However, this dependence is not
negligible either. For example, the rate of charged current
interactions of low energy neutrinos in the detector is inversely
proportional to $E_\mu$, as it can easily seen by using
Eq.~(\ref{numeric}).

\begin{figure}
\begin{center}
\epsfig{file=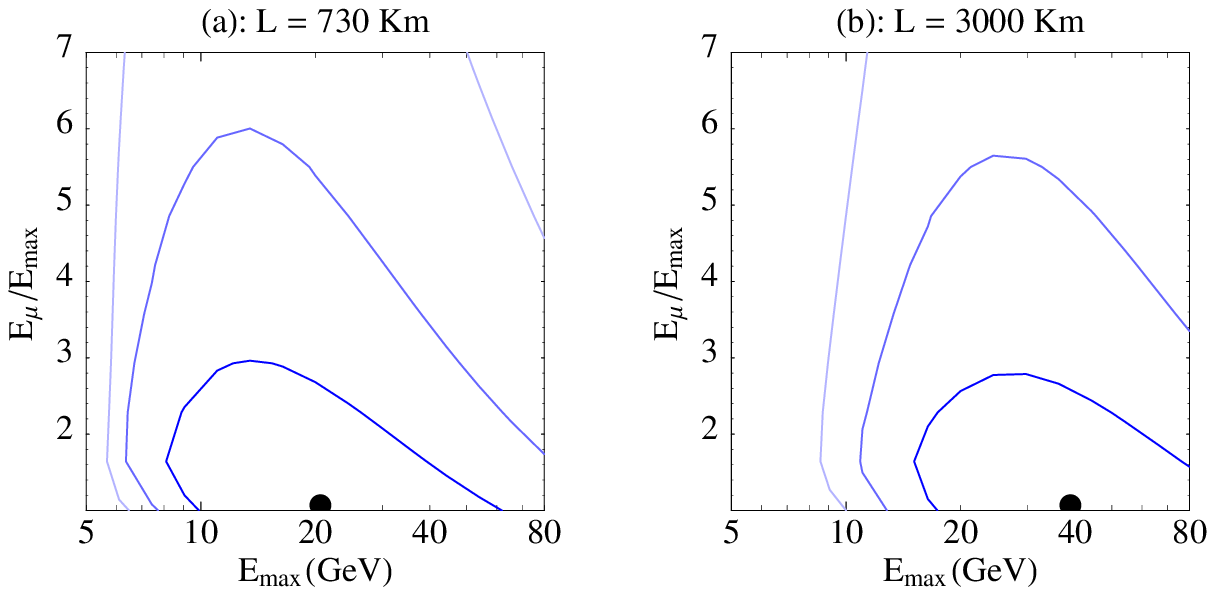,width=0.85\textwidth}
\end{center}
\mycaption{Optimization of $\emax$ and $E_\mu/\emax$ in the case of
a long baseline of $L=732\,\text{km}$, (a), and of a very long
baseline of $L=3000\,\text{km}$, (b). The black spots indicate the
values of $\emax$ and $E_\mu/\emax$ that give the best relative
statistical error. The three contour lines surround the regions where
the relative statystical error is 80\%, 60\%, 40\% of the optimal
value respectively.}
\label{fig:opt}
\end{figure}

The optimization of $\emax$ and $E_\mu$ is shown in Fig.~\ref{fig:opt}
for the two baselines we consider, $L=732\,\text{km}$
(Fig.~\ref{fig:opt}a) and $L=3000\,\text{km}$ (Fig.~\ref{fig:opt}b).
The values of $\emax$ and $E_\mu$ that maximize $|\aCP/\dast|$ are
shown by the black spots in the $E_\mu$--$E_\mu/\emin$ planes of
Fig.~\ref{fig:opt}. The three contour lines around the spots surround
the regions where $|\aCP/\dast|$ is larger than 80\%, 60\%, 40\% of the
optimal value respectively. The figures show that there is a certain
freedom in choosing the values of $\emax$ and $E_\mu$. In what follows
we will use the optimal values $\emax=E_\mu=20\GeV$ in the $L=732\,\text{km}$ case and
$\emax=E_\mu=40\GeV$ in the $L=3000\,\text{km}$ one. However, using
e.g.\ $\emax=20\GeV$ and $E_\mu=50\GeV$ for $L=732\,\text{km}$ would
involve a reduction of $|\aCP/\dast|$ of 20\% only. The dependence of
these results on the neutrino parameters is negligible.
\medskip

Having described the values of $\Nup/\Num$ $E_\mu$, $\emin$, $\emax$
we will use, we now move to the discussion of what amount of parameter
space could be covered by using a total number of $N_\mu=\Nup+\Num$
useful muon decays and a $\NKT$ kT detector. 
First of all we describe the parameter space. The neutrino
parameters are $\theta_{12}$, $\theta_{23}$, $\theta_{13}$, $\dm{21}$,
$\dm{32}$, $\delta$. The constraints on them we use are those
obtained by the standard fits of the atmospheric and solar neutrino
data~\cite{fukuda:98b,GHPV,fukuda:99a,barger:98a} taking
into account the constraints given by the CHOOZ
experiment~\cite{apollonio:97a}.  Since the uncertanties on $\theta_{12}$ and
$\theta_{23}$ do not significantly affect the results, we set $\sin^2
2\theta_{12}= 0.8$~\cite{GHPV} and $\sin^2 2\theta_{23}=1$. The
dependence of the results on $\delta$ is trivial since $\aCP$ is to a
good approximation linear in $\sin\delta$. This is because $\PCPV$ is
strictly linear in $\sin\delta$ and $\PCPC$ depends on $\delta$ only
through sub-leading $\dm{21}$ terms\footnote{Moreover, the
CP-conserving quantity $\PCPC$ depends on $\delta$ only through
$\cos\delta$, so that the dependence on $\delta$ is particularly mild
when the CP-violation is large.}. We are therefore led to consider a
three-dimensional parameter space described by \globallabel{constr}
\begin{gather}
10^{-3}\eV^2 \leq |\dm{32}| \leq 10^{-2}\eV^2~\cite{fukuda:98b,GHPV}
\mytag \\ 
2\cdot 10^{-5}\eV^2 \leq \dm{21} \leq 3\cdot
10^{-4}\eV^2~\cite{fukuda:99a} \mytag \\ 
\theta_{13} \leq \text{CHOOZ and 3-$\nu$ fit
limits~\cite{barger:98a}.} \mytag 
\end{gather}
Since the CP-violating effects crucially depend on $\dm{21}$, we plot
our results in the $\sin^2 2\theta_{13}$--$\dm{21}$ plane for different
values of $|\dm{32}|$. We first show and discuss the results for the
$L=732\,\text{km}$ baseline.

\subsection{The $L=732\,\text{km}$ baseline}
\label{subsec:732}

\begin{figure}
\begin{center}
\epsfig{file=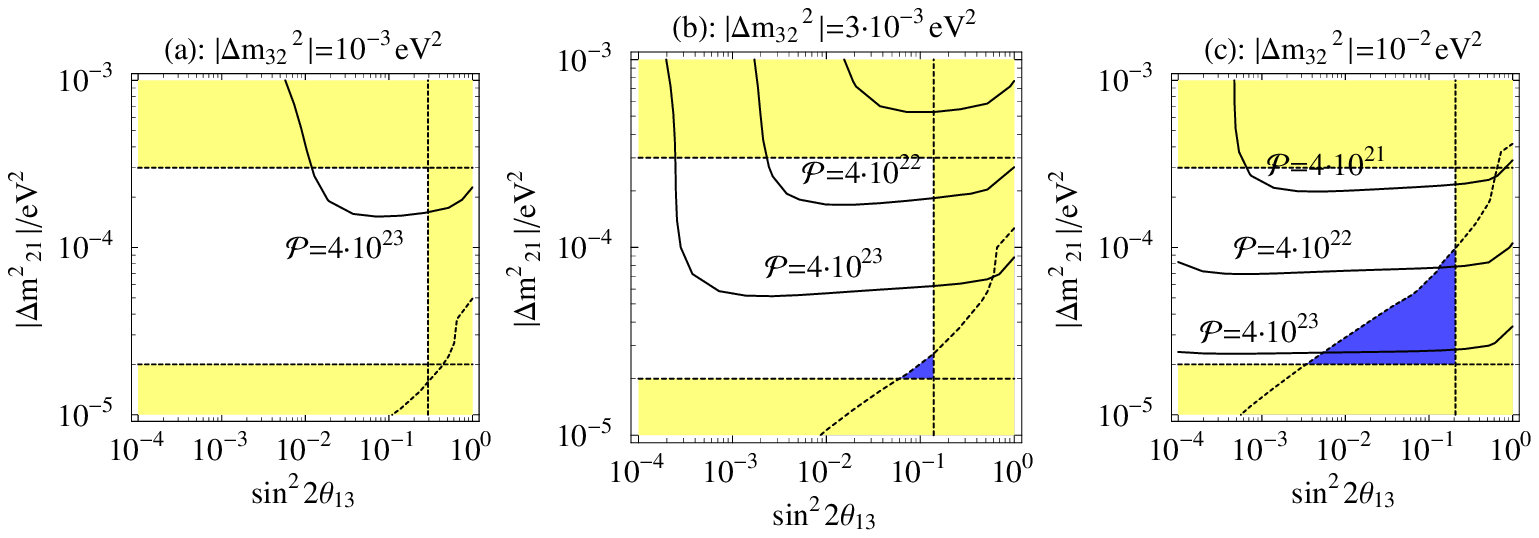,width=\textwidth}
\end{center}
\mycaption{Capability of a neutrino factory using a baseline of
$L=732\,\text{km}$ in the $\sin^2 2\theta_{13}$--$\dm{21}$ plane for
$|\dm{32}|=10^{-3}\eV^2$, (a), $|\dm{32}|=3\cdot 10^{-3}\eV^2$, (b),
and $|\dm{32}|=10^{-2}\eV^2$, (c). The rectangular windows inside the
light-shadowed areas correspond to the constraints~(\ref{constr}). In
the dark-shadowed regions the systematic error on $\aCP$ due to matter
effects exceeds $|\aCP|$. The size of these regions depends on the
precision of the future determination of $|\dm{32}|$ and
$\theta_{13}$. In order to cover the portion of parameter space above
the solid line corresponding to a given value of ${\cal P}$, values of
$N_\mu$, $\NKT$ and $\epsilon_{\mu^\pm}$ such that
$N_\mu\NKT/(n_\sigma/5)^2(\epsilon_{\mu^\pm}/30\%)\sin^2\delta >
{\cal P}$ should be used.}
\label{fig:par1}
\end{figure}

\noindent
In Fig.~\ref{fig:par1} we show the portion of the $\sin^2
2\theta_{13}$--$\dm{21}$ parameter space where an evidence of
CP-violation could be obtained for three different values of
$|\dm{32}|$: the present central value, $|\dm{32}|=3\cdot
10^{-3}\eV^2$ (Fig.~\ref{fig:par1}b in the center), and two values at
the borders of the presently allowed region, $|\dm{32}|=10^{-3}\eV^2$
(Fig.~\ref{fig:par1}a) and $|\dm{32}|=10^{-2}\eV^2$
(Fig.~\ref{fig:par1}c).  The rectangular windows inside the light
shadowed areas represent the part of parameter space allowed by the
constraints~(\ref{constr}).  The dark shadowed parts of the plots
represent the portion of parameter space where an evidence of
CP-violation could not be obtained because the systematic error
exceeds the asymmetry itself even for maximal CP-violation. We see
that the systematic error due to matter effects does not represent a
serious problem for the $L=732\,\text{km}$ baseline, especially when
$|\dm{32}|$ is low. Note that the size of this systematic error
depends on the precision of the future determinations of $|\dm{32}|$
and $\theta_{13}$. If precisions higher than those assumed here could
be achieved, the dark-shadowed regions in Fig.~\ref{fig:par1} would be
smaller. Fig.~\ref{fig:par1} shows that the relative systematic error
$|\dasy/\aCP|$ grows with $|\dm{32}|$. This is because $\aCP$ is
almost independent of $\dm{32}$ when $\theta_{13}$ is not too
small~\cite{DFLR}, whereas $|\dasy|$, roughly proportional to
$|a_{\text{m}}|$, grows with $|\dm{32}|$, as illustrated by
Eq.~(\ref{amapp}).

The region of the parameter space where the statistics is sufficient
to exclude a fluctuation with a CL corresponding to $n_\sigma$
standard deviations is shown in the plots of Fig.~\ref{fig:par1} for
different values of the number of useful muon decays $N_\mu$, of the
size of the detector in kT, $\NKT$, and of the size of CP-violation,
$|\sin\delta|$ (regions above the solid lines). Since $\aCP\simeq
\aCPd\cdot \sin\delta$ and since
$\dast\propto 1/\sqrt{N_\mu\NKT}$, the regions shown actually
depend on the combination
\begin{equation}
\label{calP}
{\cal P} =
\frac{N_\mu\NKT}{(n_\sigma/5)^2}\cdot
\frac{\epsilon_{\mu^\pm}}{30\%}\cdot \sin^2\delta
\end{equation}
only.  For a 30\% efficiency, maximal CP-violation and a CL
corresponding to the reference value $n_\sigma=5$, ${\cal P}$ is
simply the product of $N_\mu$ and $\NKT$. The three solid lines in
Figs.~\ref{fig:par1} (one in Fig.~\ref{fig:par1}a) correspond to three
possible values of $\cal P$: the smallest one, ${\cal P}=4\cdot
10^{21}$, corresponds (for $\sin^2\delta=1$ and $n_\sigma=5$) to the
reference set-up considered in~\cite{derujula:98a}, a 10 kT detector
and $2\cdot 10^{20}$ useful muon decays in both the CP-conjugated
channels. The intermediate value, ${\cal P}=4\cdot 10^{22}$,
corresponds e.g.\ to an improvement of one order of magnitude in the
intensity of the muon source. The highest value, ${\cal P}=4\cdot
10^{23}$, requires a further improvement, e.g.\ a 30 kT detector with
very high efficiency. Some of these options are currently under
discussion~\cite{lyon}.

Fig.~\ref{fig:par1} clearly shows that the sensitivity lines strongly
depend on the value of $|\dm{32}|$. This is because the statistics,
proportional to $\sin^2\Delta_{32}$, grows with $|\dm{32}|$, whereas
the asymmetry is almost independent of $\dm{32}$ for $\theta_{13}$ not
too small~\cite{DFLR}. A $|\dm{32}|$ value around the present central
value would allow to cover a major part of the parameter space with
the highest value of ${\cal P}$. If $|\dm{32}|$ were at the top of the
presently allowed range, that value would allow to cover essentially
all the parameter space, while the intermediate value of ${\cal P}$ in
this case would be enough to cover a good half of it. At the bottom of
the $|\dm{32}|$ range getting an evidence of leptonic CP-violation
would only be possible in a small part of the parameter space even
with the highest neutrino fluxes, unless a longer baseline is
used. This possibility will be considered in the next
Subsection.

Values of $\dm{21}$ above the range~(\ref{constr}b) are allowed in
``non-standard'' analysis of solar neutrino data where the possibility
of an unknown systematic uncertainty in one of the solar experiments is
taken into consideration~\cite{barbieri:98c}. Fig.~\ref{fig:par1}
shows that relatively low values of ${\cal P}$ could cover the
corresponding region.

\begin{figure}
\begin{center}
\epsfig{file=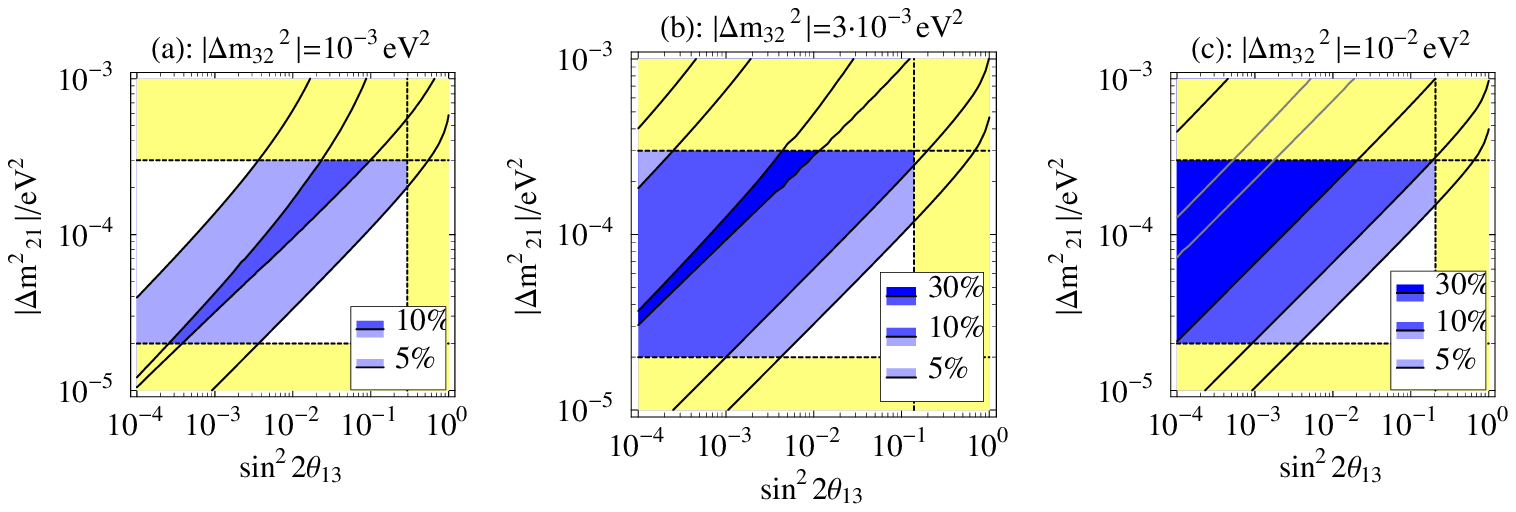,width=\textwidth}
\end{center}
\mycaption{Contour lines of $|\aCPd|\simeq|\aCP/\sin\delta|$ in the
$\sin^2 2\theta_{13}$--$\dm{21}$ plane for $|\dm{32}|=10^{-3}\eV^2$,
(a), $|\dm{32}|=3\cdot 10^{-3}\eV^2$, (b), and
$|\dm{32}|=10^{-2}\eV^2$, (c), in the case of a $L=732\,\text{km}$
baseline. In (c) the contour lines inside the darker region correspond
to a 70\% asymmetry.}
\label{fig:asy1}
\end{figure}

For completeness, we show in Fig.~\ref{fig:asy1} three contour plots of
$|\aCPd|\simeq|\aCP/\sin\delta|$ in the same sections of the
three-dimensional 
parameter space as in Fig.~\ref{fig:par1}. The asymmetry values
corresponding to the contour lines are specified in the legends. The
structure of the plots is easily explained: as far as the $\dm{21}$
effects in the CP-conserving part of the oscillation probability can
be neglected,
\begin{equation}
\label{expl}
\PCPV\propto\sin 2\theta_{13}\qquad\text{and} \qquad 
\PCPC\propto\sin^2 2\theta_{13},
\end{equation}
so that the CP-asymmetry gets larger when $\theta_{13}$ gets
smaller, $\aCP\propto 1/\sin 2\theta_{13}$~\cite{DFLR}. However, at
some point the smallness of $\theta_{13}$ makes the $\dm{21}$ effects
in $\PCPC$ important so that $\aCP$ can vanish when
$\theta_{13}\rightarrow 0$ as it should.

Large asymmetries are therefore associated with small $\theta_{13}$
and with small oscillation rates. One can then wonder whether having a
large asymmetry would not represent a disadvantage from the point of
view of statistics. Nevertheless, this is not the
case. From~(\ref{expl}) we see in fact that, as far as we can set
$\dm{21}=0$ in $\PCPC$, $\aCP/\dast\propto \PCPV/\sqrt{\PCPC}$ is
actually independent of $\theta_{13}$. This explains why the
sensitivity lines in Figs.~\ref{fig:par1} are approximately horizontal
as far as $\theta_{13}$ is not too small.

\subsection{Very long baselines}
\label{subsec:3000}

\noindent
We have seen in the previous Subsection that the possibility of
covering a large region of the parameter space with a
$L=732\,\text{km}$ baseline crucially depends on $|\dm{32}|$. A better
determination of $|\dm{32}|$ will come in the next years from the
upcoming long-baseline experiments. If $|\dm{32}|$ turned out to be
close or above the present central value and if the fluxes under
discussion could be achieved, the $L=732\,\text{km}$ baseline could be
enough. If, on the contrary, $|\dm{32}|$ turned out to be low or the
CP-violation size $|\sin^2\delta|$ were small a longer baseline would
be necessary.  The reason why a longer baseline helps from the point
of view of overcoming the statistical error is that the asymmetry
grows with $L$, whereas the statistics, and therefore $\dast$, is
approximately independent of $L$. This holds however only as far as
$|\Delta_{32}|\ll\pi/2$, or $L\ll400\,\text{km}\, E_\nu(\GeV) (3\cdot
10^{-3}\eV^2/|\dm{32}|)$.

\begin{figure}
\begin{center}
\epsfig{file=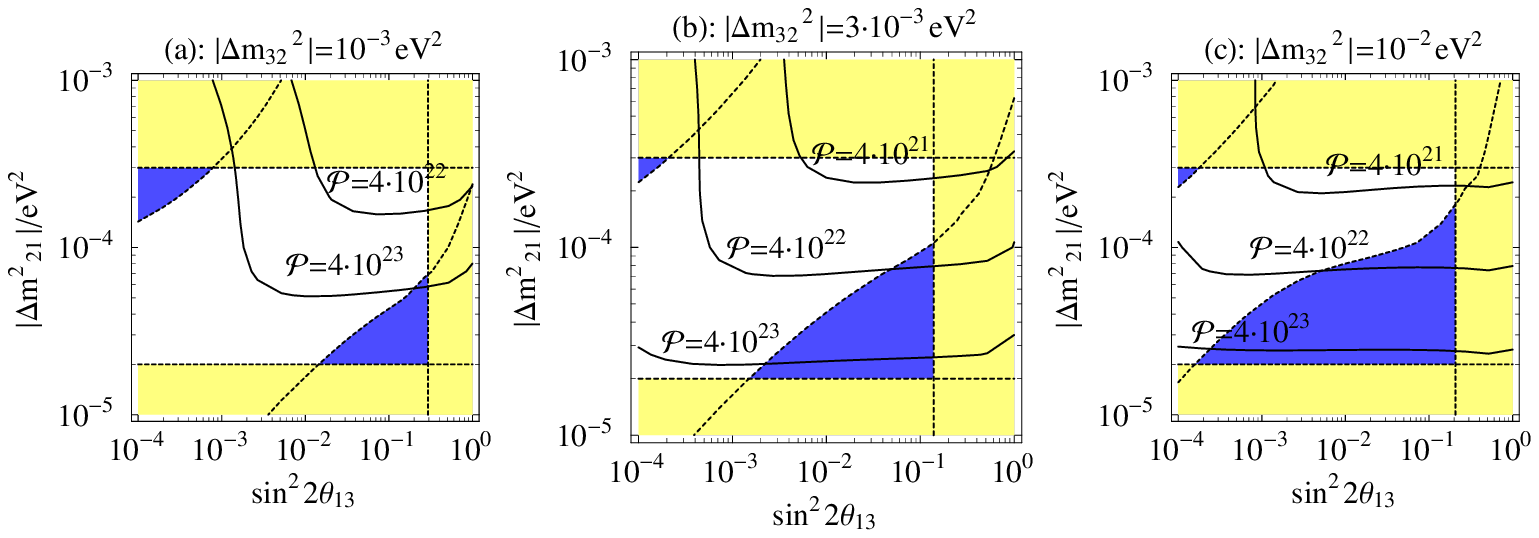,width=\textwidth}
\end{center}
\mycaption{Same as in Fig.~\ref{fig:par1} for a baseline
of $L=3000\,\text{km}$.}
\label{fig:par2}
\end{figure}

In this Subsection we consider the possibility of a
$L=3000\,\text{km}$ baseline. One can wonder whether having an even
longer baseline could not to be better. However, this is not
necessarily the case. First of all, $\sin^2\Delta_{32}$ reaches its
maximum when $L\sim 400 \,\text{km}\, E_\nu(\GeV) (3\cdot
10^{-3}\eV^2/|\dm{32}|)$, so that for longer baselines the statistics
decreases significantly. Moreover, matter effects grow with $L$ faster
than the asymmetry, as pointed out in Section~\ref{subsec:errors}.
This is also confirmed by a comparison of Fig.~\ref{fig:par1} with
Fig.~\ref{fig:par2}, that only differs from Fig.~\ref{fig:par1} for
the length of the baseline, $L=3000\,\text{km}$ instead of
$L=732\,\text{km}$. We see that the dark-shadowed regions where the
matter effect uncertainties are too large are larger in the present
$L=3000\,\text{km}$ case. Therefore, for longer baselines matter effects
would represent a major problem. Note, however, that these
uncertainties depend on the precision of the future determinations of
$|\dm{32}|$ and $\theta_{13}$ and they could therefore be smaller than
what estimated here (see Section~\ref{subsec:errors}) .

We see from Fig.~\ref{fig:par2} that using a $L=3000\,\text{km}$
baseline and the highest value of ${\cal P}$ would allow to cover
about 2/3 of the $\dm{21}$ parameter space (for $\sin^2
2\theta_{13}\gtrsim 2\cdot 10^{-3}$) in the most unfavourable case
$|\dm{32}|=10^{-3}\eV^2$ and essentially all of it for larger values
of $|\dm{32}|$. For such values, the intermediate value of ${\cal
P}$ would be enough to cover a good half of the $\dm{21}$ parameter
space.  

\begin{figure}
\begin{center}
\epsfig{file=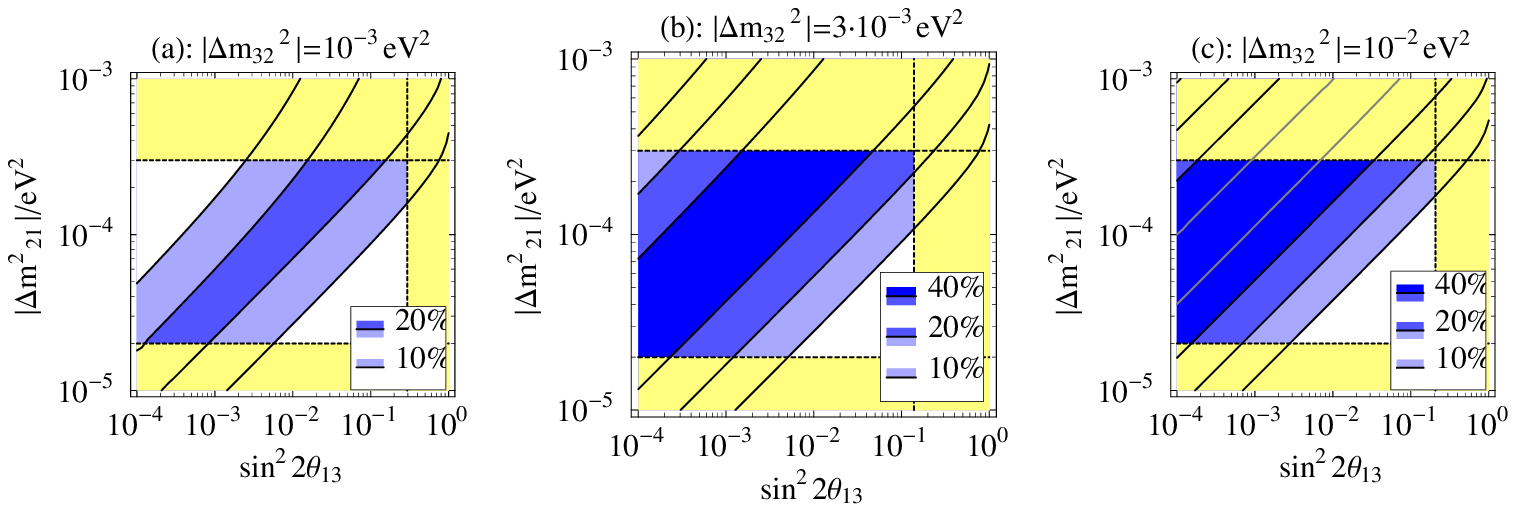,width=\textwidth}
\end{center}
\mycaption{Same as in Fig.~\ref{fig:asy1} for a baseline
of $L=3000\,\text{km}$.}
\label{fig:asy2}
\end{figure}

Finally Fig.~\ref{fig:asy2} shows the contour plot of
$|\aCPd|\simeq|\aCP/\sin\delta|$ for the
$L=3000\,\text{km}$ baseline. 

\section{Discussion and conclusions}
\label{sec:conc}

\noindent
The outcomes of the upcoming neutrino experiments will play a crucial
role in assessing whether it could be worth making an effort towards a
CP-violation measurement. First of all, they will help to understand
whether a light sterile neutrino exists and mixes significantly with
the active ones. If this were the case, the detection of leptonic
CP-violation would be within the capabilities of an
intermediate-baseline experiment~\cite{DFLR,DGHR}.  If, on the
contrary, the standard three neutrino scenario were confirmed, the
size and the possibility of measuring CP-violation would crucially
depend on the mechanism accounting for the solar neutrino deficit. If
the large mixing angle solution were ruled out, there would be no hope
of measuring leptonic CP-violation through neutrino oscillations. In
the case of the small angle solution, this can be seen using
Figs.~\ref{fig:par1} and~\ref{fig:par2}. In fact, those figures also
apply to the small angle solution provided that the parameter ${\cal
P}$ is interpreted as
\begin{equation}
\label{calP2}
{\cal P} =
\frac{N_\mu\NKT}{(n_\sigma/5)^2}\cdot\frac{\epsilon_{\mu^\pm}}{30\%}
\cdot\sin^2\delta\,\sin^2
2\theta_{12},
\end{equation}
where $\theta_{12}$ is now constrained to be within the small angle
range. Therefore, in order to get same coverage in the $\sin^2
2\theta_{13}$--$\dm{21}$ plane as in the large angle case, it would be
necessary to increase $N_\mu\cdot \NKT$ by a factor
$1/\sin^22\theta_{12} \sim 200$.\footnote{Moreover, that would not
still be enough since the $\dm{21}$ range of the small angle solution
is approximately one order of magnitude lower than the $\dm{21}$ range
of the large angle solution.} If, on the other hand, the large mixing
angle solution were preferred by the data, the measurement of a
leptonic CP-violation in the
$\nu_e\leftrightarrow\nu_\mu$/$\bar{\nu}_e\leftrightarrow\bar{\nu}_\mu$
channel would represent an exciting experimental challenge.

In the latter case, a precise assessment of the capabilities of a
neutrino factory, as well as the determination of the best
experimental configuration (essentially the baseline length) would
strongly depend on the value of two parameters: the intensity of the
muon source and the precise value of $|\dm{32}|$ which will be
provided by the long-baseline experiments.  Figs.~\ref{fig:par1}
and~\ref{fig:par2} show what portion of parameter space could be
covered for different values of these two parameters and for two
possible baselines, $L=732\,\text{km}$ (Fig.~\ref{fig:par1}) and
$L=3000\,\text{km}$ (Fig.~\ref{fig:par2}). If $|\dm{32}|$ turned out to
be in the upper part of its present range, $|\dm{32}|\sim
10^{-2}\eV^2$, and very high-intensity sources could be achieved, the
$L=732\,\text{km}$ baseline could be enough to cover a relevant part
of the parameter space. On the other hand, if $|\dm{32}|$ turned out
to be lower, using an even longer baseline, $L\sim 3000\,\text{km}$,
could be necessary. Using baselines longer than $3000\,\text{km}$
would require knowing matter effects with a higher precision than that
assumed in this paper.  Note that, since the relative systematic error
due to matter effects grows both with $L$ and $|\dm{32}|$ (as opposed
to the statistical one, that decreases both with $L$ and $|\dm{32}|$)
the very long baseline option could turn out to be less appropriate
for the higher values of $|\dm{32}|$.

Finally, a neutrino factory would give the possibility of measuring
$\theta_{13}$ with a high accuracy. This is not only due to the pure
and high-intensity neutrino flux, but also due to the possibility of
measuring both the $\nu_e\leftrightarrow\nu_\mu$ and
$\bar{\nu}_e\leftrightarrow\bar{\nu}_\mu$ channels. A determination of
$\sin^2 2\theta_{13}$ made by measuring a single oscillation
probability only could in fact be affected by a systematic uncertainty
up to 30\% associated with CP-violation effects~\cite{DFLR} (see also
Figs.~\ref{fig:asy1} and~\ref{fig:asy2}). On the contrary, measuring
both $P_\reu$ and $P_\reub$ would allow to get rid of the CP-violating
part of the probability $\PCPV$ by computing
$(P_\reu+P_\reub)/2=\PCPC$ and to get rid of the further dependence of
$\PCPC$ on $\delta$ by computing the CP-asymmetry $\aCP$.

\subsection*{Acknowledgments}

\noindent
We thank M. Freund, M.B. Gavela, J. G\'omez-Cadenas, P. Hernandez,
M. Lindner and S. Petcov for useful discussions. Work supported by the
TMR Network under the EEC Contract No.~ERBFMRX--CT960090.

\end{document}